\documentclass{optica-article}

\journal{opticajournal} 

\articletype{Research Article}

\usepackage{lineno}

\begin{document}

\title{Label-free microscope for rheological imaging of cells}

\author{Nicolas P. Mauranyapin,\authormark{1} Marino Lara Alva,\authormark{1}, Daniel Yan,\authormark{2}, Zhe Yang,\authormark{3}, Jackson D. Lucas\authormark{1}, Alex Terrasson\authormark{1}, Michael A. Taylor,\authormark{4} Rohan Teasdale,\authormark{3} Yun Chen\authormark{2} and Warwick P. Bowen\authormark{1,*}}

\address{\authormark{1} Australian Research Council Centre of Excellence in Quantum Biotechnology (QUBIC), The University of Queensland, Australia \\
\authormark{2} Department of Mechanical Engineering,
Center for Cell Dynamics, Institute for NanoBioTechnology (INBT)
Johns Hopkins University, Baltimore, USA\\
\authormark{3} School of Biomedical Sciences, The University of Queensland, Australia.\\
\authormark{4} University of Otago, Dunedin, New Zealand}

\email{\authormark{*}w.bowen@uq.edu.au} 

\begin{abstract*} 
Many essential cellular functions depend on the viscoelastic properties of the cytoplasm. While techniques such as optical tweezers and atomic force microscopy can measure these properties, their reliance on localized probes prevents intracellular imaging and perturbs native cellular behaviour. Label-free microscopy offers non-invasive alternatives that are capable of imaging.
However, bandwidth limitations often confine these techniques to the assessment of static mechanical properties or to measurements at gigahertz frequencies, which both lie outside the interesting frequency range typically associated with cellular viscoelasticity.
Here, we introduce a label-free microscope capable of imaging the viscoelastic properties of cells at frequencies relevant to biology. The microscope measures intracellular viscoelasticity -- twenty times faster than previous label-free approaches -- and does this with diffraction limited resolution. The measurements reveal characteristic viscoelastic features that were previously inaccessible, allowing quantitative rheology of the cellular cytoskeleton. 
We apply the microscope to live cancer cells.
The rheological images produced identify spatial variations in cellular mechanics, allow active and thermal processes to be distinguished pixel-by-pixel, and enable the state of the cell to be visualised over time and in the presence of stress. The microscope is also able to resolve cellular structures that are invisible to regular phase-sensitive imaging, and do this with high contrast. The ability to image both intracellular viscoelasticity and activity offers a powerful tool to advance fundamental cell biology, cancer research, clinical diagnostics, and drug development.
\end{abstract*}

\vspace{10 mm}

Cell function and health depend critically on the physical properties of the cytoplasm. In particular, cytoplasmic viscoelasticity and dynamics govern essential processes such as intracellular transport, cell motility, and mechanochemical transduction~\cite{mofrad2009rheology}. Alterations in these properties are associated with chronic inflammation, neurodegenerative diseases, and cancer~\cite{mofrad2009rheology}.
Microrheological techniques have been developed to probe cytoplasmic viscoelasticity by tracking the motion of tracer particles such as micro-beads \cite{gittes1997microscopic}, magnetic rods \cite{chevry2013intracellular}, fluorescent tags~\cite{kole2004intracellular} or AFM tips \cite{rigato2017high}, either embedded within the cell or attached to its membrane. These methods have revealed that cells exhibit a complex shear modulus, reflecting both elastic and viscous behaviour dependent on excitation frequency ~\cite{rigato2017high,chaubet2020dynamic,hoffman2006consensus}.
However, the probe particles can disrupt cellular function or even lead to cell death~\cite{wu2012high}. More critically, particle tracking is incompatible with continuous, full-field spatial imaging.

Label-free microscopy techniques are well suited for imaging applications and provide non-invasive approaches to probing intracellular mechanics. Among these, Brillouin microscopy enables mapping of the viscoelastic properties of biological tissues and cells at the micrometre scale by measuring gigahertz-frequency shifts arising from the interaction between incident light and thermally excited acoustic phonons within the sample, which are linked to the sample’s mechanical response. This approach has enabled the investigation of intracellular biomechanics in living cells \cite{scarcelli2015noncontact} and shows promise for early diagnosis of diseases such as cancer and keratoconus \cite{prevedel2019brillouin}. However, the narrow gigahertz frequency window accessed by Brillouin microscopy lies well outside the frequency range typically associated with many biologically relevant mechanical processes.
Alternatively, techniques such as Rotating Coherent Scattering (ROCS)~\cite{junger2022100} and Interferometric Scattering (iSCAT)~\cite{hsiao2022spinning,trelin2024chiscat} have demonstrated exceptional sensitivity for detecting the motion of endogenous cellular components without labels, enabling the observation of single-protein dynamics in solution~\cite{young2018quantitative} and intracellular dynamics in living cells~\cite{hsiao2024probing}. Yet, these approaches have not been extended to quantitative rheology, owing both to their limited effective bandwidth, which hinders resolution of key rheological features~\cite{rigato2017high}, and to the incomplete understanding of the biophysical origin of the detected signals.
Conversely, Dynamic Light Scattering (DLS)~\cite{joo2010diffusive} and Coherent Bright-Field (COBRI) microscopy~\cite{huang2017coherent} provide access to high measurement bandwidths. But, their high background noise and limited sensitivity have so far precluded applications to subcellular microrheology, restricting measurements to discrete tracer particles and preventing continuous, label-free imaging. Although microrheology and imaging have been demonstrated using DLS, these implementations rely on ensemble averaging over length scales of tens of micrometres~\cite{krajina2017dynamic,postnov2020dynamic}, thereby precluding subcellular-resolution mapping, which is essential for resolving cellular morphology and active processes.

Here, we introduce rheoSCAT microscopy, a label-free technique capable of subcellular rheological imaging of cells. The microscope achieves this by greatly suppressing laser and technical noise, so that minute phase fluctuations driven by nanoscale endogenous motion can be observed. This allows measurements of intracellular dynamics with 50~kHz bandwidth, twenty times higher than previous techniques~\cite{hsiao2024probing}, and with sub-micron spatial resolution. Crucially, the microscope resolves high frequency dynamical features that were previously inaccessible. Modelling and reference experiments with optical tweezers allow us to connect these features to key properties of the cytoplasm, namely the crossover frequencies between viscous and elastic behaviour, and between active and thermal forces.
This new understanding enables quantitative rheological imaging, which
we prove with experiments on living cancer cells. We show that the microscope can image spatial variations in viscoelasticity and cellular activity, separately resolve active and thermal motion, and spatially track how cells respond to stress over time when exposed to chemical assault. Beyond rheology, we  demonstrate that imaging dynamical fluctuations provides improved contrast compared to static phase-sensitive imaging, resolving structures on the surface of cells that were otherwise unresolvable.

The ability of rheoSCAT microscopy to quantitatively image cellular dynamics and microrheology could shed new light on basic biological processes that remain poorly understood, such as intracellular transport~\cite{li2018intracellular}, energy transfer~\cite{dzeja2003phosphotransfer}, and the mechanics of cellular signalling~\cite{kamata1999redox}. It could also have applications in medical research, contributing to the development of new diagnostic methods, drug discovery, the study of neurodegenerative diseases, and regenerative medicine.

\section*{Results}

\subsection*{rheoSCAT microscopy}

\begin{figure}[ht!]
\centering\includegraphics[width=\textwidth]{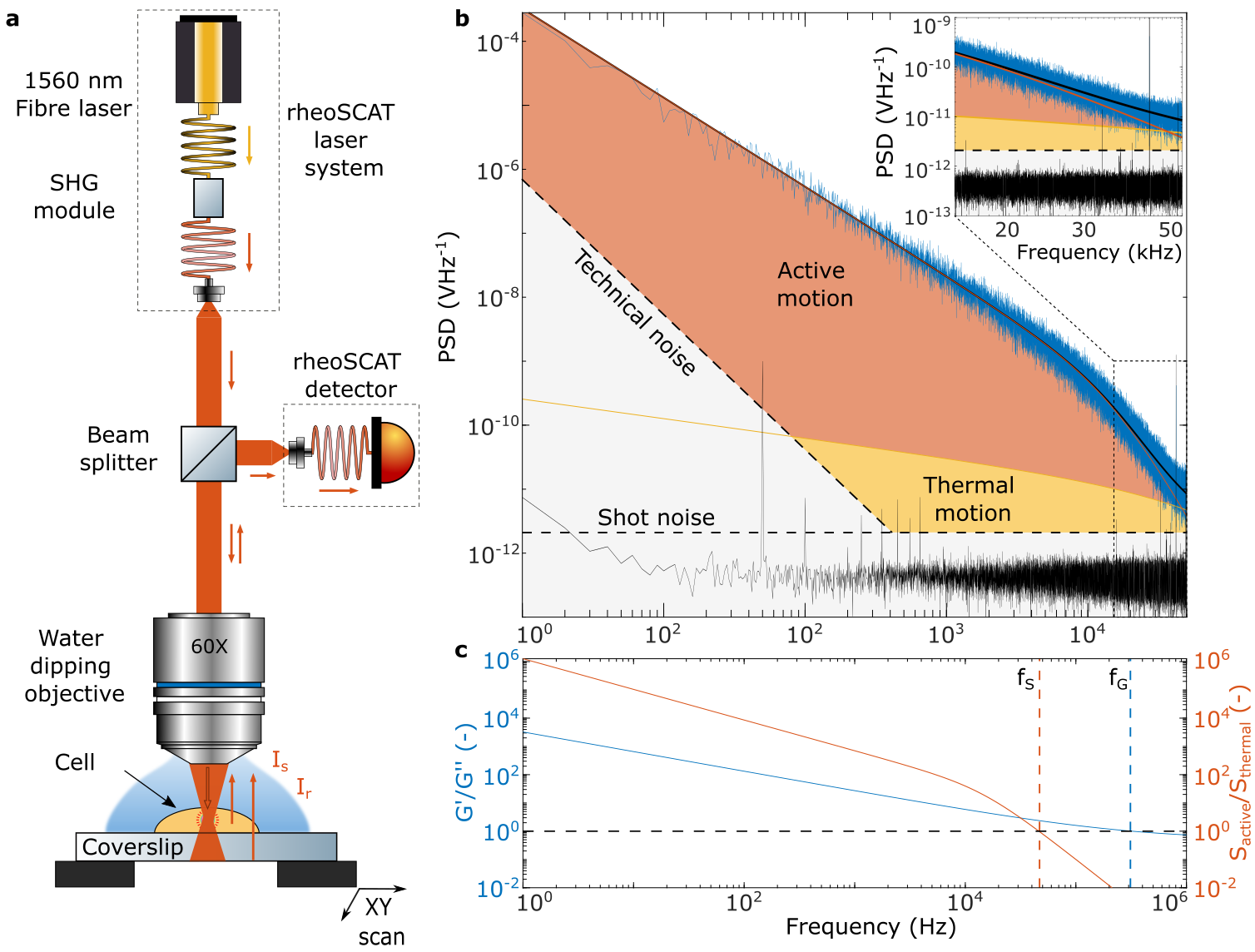}
\caption{\textbf{Experimental setup}. \textbf{a} Schematic of the rheoSCAT microscope. \textbf{b} Power spectral density (PSD) of the photocurrent time traces (blue) inside a HeLa cell averaged over 10 measurements. Curve fitting the blue PSD using the model $S(f)$ is displayed in black and it thermal and active motion contributions are displayed in yellow and red with yellow and red shadings respectively. The horizontal dashed lines represent the averaged shot noise on the measurement and the diagonal dashed line represent the technical noise of the microscope. The black trace represents the electronic noise of the photodetector. Inset: zoom over the high frequencies. \textbf{c} Blue: ratio of the elastic ($G'$) and viscous ($G''$) modulus versus frequency. Red: ratio of the active $S_{active}$ and thermal $S_{thermal}$ force spectrum versus frequency. These ratio are calculated from the PSD fit in \textbf{b}. The blue and red dashed line represents crossover frequencies $f_G$ and $f_S$ where $G'/G'' =1$ and $S_{active}/S_{thermal}=1$ respectively.}
\label{fig:setup}
\end{figure}

Figure~\ref{fig:setup}\textbf{a} provides a schematic of the rheoSCAT microscope which, like iSCAT~\cite{piliarik2014direct}, harnesses the interference between the scattered light from a specimen and the reflected light from a microscope cover slip that it is attached to. Upon detection, the interference creates a photocurrent which can be approximated to first order as (see Supplementary Information)
\begin{equation}
        i(t) \propto \sqrt{I_{\rm{r}}} \left[ 
        {\underbrace{\sqrt{I_{\rm{r}}} + 2\sqrt{ I_{\rm{s}}} \rm{cos}\phi}_{\text{static phase signal}}}
        + {\underbrace{\sqrt{I_s}\delta S(t)}_{\text{bioactivity signal}}}
        + {\underbrace{2\delta N(t)}_{\text{light noise}}}
        \right] 
        + {\underbrace{\delta E(t).}_{\text{electronic noise}}}
        \label{eq:I}
\end{equation}
Here $I_r$ and $I_s$ are the mean intensities of the reflected and scattered fields, respectively, $\phi$ is the mean phase difference between the two fields, and we assume $I_r \gg I_s$, an approximation that is supported by our experimental measurements. The first term in Eq.~(\ref{eq:I}) is a static phase signal analogous to the signal from iSCAT~\cite{piliarik2014direct}. The other terms are associated with fluctuations. $\delta S(t)$ is the rheoSCAT signal due to the probed cellular constituents motion that act as endogenous microrheological probes, $\delta N(t)$ is the optical noise on the measurement, and $\delta E(t)$ is the electronic noise of the detection system.

Since the amplitude of cellular fluctuations decreases as a power law with frequency~\cite{guo2014probing}, strong noise suppression of both the electronic and light noise is required to observe rheological signals at high frequencies.

In a key innovation to suppress light noise, we develop an ultra-low noise microscope illumination source at a wavelength of 780~nm. This is based on a single frequency telecom-wavelength fibre laser with relative intensity noise lower than -135 dBc/Hz (NKT: Koheras ADJUSTIK X15, 1560~nm), this corresponds to a reduction of amplitude noise by an order of magnitude compared to typical iSCAT and other types of phase contrast microscopy illumination sources~\cite{piliarik2014direct}, allowing fainter signals to be detected. The 1560~nm light is then frequency doubled to 780~nm using a periodically poled nonlinear waveguide, preserving the low-noise characteristics of the original fibre laser while allowing operation in the biologically safe first near-infrared window.

The near-identical optical paths traversed by the reflected and signal fields (see Fig.~\ref{fig:setup}\textbf{a}), with a path difference of around $300$~µm, greatly suppresses phase fluctuations generated within the microscope compared to traditional phase-contrast or DIC microscopes~\cite{nishiwaki2019interference}. In addition, as shown in Eq.~(\ref{eq:I}), the interference amplifies the rheoSCAT signal by a factor of $\sqrt{I_{\rm r}I_{\rm s}}$ above the electronic noise floor of the detection system~\cite{mauranyapin2017evanescent}. This  suppresses the relative contribution of the electronic noise so that detection limited by technical light-noise and shot-noise is possible using a high bandwidth photoreceiver rather than a lower noise, lower bandwidth alternative such as a CMOS camera~\cite{piliarik2014direct}. Beyond improved bandwidth, the use of a photoreceiver also allows higher illumination intensities, increasing the signal-to-noise of the microscope.
The laser is focused on the sample using a water-dipping objective with 60X magnification and a numerical aperture of 1.0. Using a water dipping objective reduces the number of optical interfaces that the light must traverse, minimising spurious interference and maximising detection efficiency. The output field is collected through the same objective and spatially filtered through a single-mode optical fibre to capture only the scattered and reflected fields from the laser’s focal point, following a confocal approach~\cite{dabbs1992single}. This suppresses noise from out-of-focus light and higher-order spatial modes.

We detect the spatially filtered output field of the microscope with a 200~kHz bandwidth photodetector that can handle optical powers as high as 10~mW, far above the maximum that can typically be used with a CMOS camera. The photodetector (Newport model 2001-FS-M) has noise-equivalent power of 0.25~pW$/\sqrt{\rm{Hz}}$, allowing shot-noise limited performance even at detected powers as low as 110~pW.

\subsection*{Label-free microrheology at high bandwidth}

In a first set of experiments, we use the rheoSCAT microscope to probe the motion of cellular constituents of a cervical cancer HeLa cell. HeLa cells are cultured in an incubator for 48 hours, followed by plating on a 20~mm$\times$20~mm coverslip for an additional 48 hours, achieving approximately 50\% confluence (see Methods).

We used 13 mW of illumination power, detecting approximately 160 nW of light reflected from the coverslip and scattered by the specimen with an acquisition time of 1~s. The detected signal is analysed via Fourier transformation to obtain the power spectral density ($\mathrm{PSD}(f)$). A representative PSD, averaged over ten measurements, is shown in Fig.~\ref{fig:setup}\textbf{b} (blue). The rheoSCAT signal exceeds the electronic noise (black trace) and optical shot-noise (horizontal dashed line) by more than eight orders of magnitude at low frequencies, and dominates technical laser noise (diagonal dashed line) by more than two order of magnitude. Despite the rapid decay of activity with frequency, the signal remains above the noise sources even beyond 50~kHz, enabling both analysis of high-frequency bioactivity and fast data acquisition. The spatial resolution, limited by the focused laser spot size, is approximately 400~nm, well within the subcellular regime.

The strong signals observed within the cell stem from the motion of endogenous scatterers in the cellular cytoplasm. They likely originate from correlated motion of proteins and proteins complexes which arise due to the correlated nature of forces from the cytoskeleton \cite{levine2000one}. Our modelling shows that these correlated motions can be expected to occur over length scales comparable to our microscope voxel and are greatly amplified in the rheoSCAT signal relative to uncorrelated motions, scaling as the square of the number of scatterers in the microscope voxel (see Supplementary Information). Larger constituents, such as mitochondria, while individually scattering more light, are predicted to collectively contribute far less to the measured signal (see Supplementary Information).
We confirm that the observed signals are not attributable to laser noise through separate measurements on a sample without cell, which show significant lower fluctuations with a different spectral shape (diagonal dashed line in Fig.~\ref{fig:setup}\textbf{b}; see also Supplementary Information). Although our measurements do not spatially resolve the microtubules and actin filaments that form the cytoskeletal structure, they enable microrheological analysis of their collective mechanical properties, as demonstrated in previous probe-based studies \cite{gittes1997microscopic}.

To verify that rheoSCAT measurements probe cellular microrheology, we performed control experiments using 1~$\mu$m silica beads embedded in HeLa cells and tracked their motion with optical tweezers, analogous to probe-based microrheology (see Supplementary Information). The features in the resulting power spectral densities closely matched those obtained with rheoSCAT.

We employed the biophysical theoretical model for the cytoskeletal power spectral density from Ref.~\cite{ahmed2018active} to fit our experimental PSDs. The theoretical spectra (black line in Fig.~\ref{fig:setup}\textbf{b}), although originally developed for probe-based microrheology, shows excellent agreement with our measurements. Following Ref.~\cite{ahmed2018active}, the total model spectrum $S(f)$ can be decomposed into  thermal and active components as
\begin{equation}
    S(f) = \frac{S_\mathrm{thermal}(f)+S_\mathrm{active}(f)}{(6\pi r)^2|G^{*}(f)|^2}.
\end{equation}
Here $S_\mathrm{thermal}(f)= 2 k_B T\kappa\xi(f\xi)^{\alpha-1}\sin{(\pi\alpha/2)}$ is the thermal force spectrum creating Brownian motion, $k_B$ is the Boltzmann constant, $T=300$~K the ambient lab temperature,  $\kappa$ represents the elastic caging force from the cytoskeleton reflecting how strongly cell constituents are confined, $\xi$ the passive relaxation time which characterizes the viscoelastic response of the cytoplasm and $\alpha$ is the cytoplasm shear modulus $G^*(f)$ power law exponent. $\alpha$ can also be related to the motion diffusion with $\alpha=1$ indicating normal diffusion, $\alpha<1$ sub-diffusion and $\alpha>1$ super-diffusion~\cite{gittes1997microscopic}.
$S_\mathrm{active}(f)=(2 k_B T_a\kappa\xi)/((f\xi)^{2\alpha}(1+(f\tau)^2)) $ is the active force spectrum arising from intracellular processes such as active diffusion~\cite{almonacid2015active} and molecular motor activity~\cite{fletcher2010cell}, with $T_a$ the effective active temperature representing the strength of active fluctuations and $\tau$ the active persistence time representing the correlation timescale of active forces. $r$ is the radius of the probe particle and $|G^{*}(f)|^2 = G'^2+G''^2= (\kappa/6\pi r)^2\left[(f\xi)^{2\alpha}+2(f\xi)^{\alpha}\cos{(\pi\alpha/2})+1\right]$ with $G'(f)$ and $G''(f)$ the elastic and viscous modulus of the cytoplasm respectively.

Employing the above model allows us to separate the thermal and active motion in the measured signal. The contributions are displayed in Fig.~\ref{fig:setup}\textbf{b} in yellow and red shading. As expected for living cells, the active motion dominates over the thermal motion at low frequencies below the crossover frequency $f_S=46$~kHz and thermal noise dominates at high frequencies.

From the fit of $S(f)$ to the PSDs (black line in Fig.~\ref{fig:setup}\textbf{b}) we can extract values for the fitting parameters $\kappa$, $T_a$, $\xi$, $\alpha$ and $\tau$. The value of $\kappa=(3.6 \pm 0.1 ) \times 10^8$~a.u. is uncalibrated because the properties and number of scatterers are not known precisely.
The remaining parameters, however, provide quantitative insights into intracellular mechanics. For the PSD in Fig.~\ref{fig:setup}\textbf{b}, we find that $T_a/T = 5 \pm 1$, $\xi = 10 \pm 6 $~ms, $\alpha=0.69 \pm 0.006$ and $\tau = 68 \pm 4$ms where the $\pm$ uncertainties represent one standard deviation from fitting uncertainties. The values of $T_a$, $\xi$ and $\alpha$ are consistent with previously reported results from microrheology~\cite{ahmed2018active}, supporting that rheoSCAT can be used as a quantitative microrheological microscope. The active persistence time $\tau$ is around two orders of magnitude larger than values typically reported using bead-based microrheology~\cite{ahmed2018active}. This discrepancy is expected because micrometer-sized beads predominantly sample local, rapidly decorrelating active fluctuations. In contrast, our approach probes correlated endogenous motion within the cytoskeletal network, capturing collective stress dynamics that persist over time significantly longer than local motor-driven events \cite{koehler2011collective}

Since $\kappa$ is not calibrated, the elastic ($G'(f)$) and viscous ($G''(f)$) moduli cannot be expressed in absolute units, but their ratio can be determined as
\begin{equation}
\frac{G'(f)}{G''(f)} = \frac{(f\xi)^\alpha \cos{(\pi\alpha/2)} + 1}{(f\xi)^\alpha \sin{(\pi\alpha/2)}}.
\label{eq:ratioG}
\end{equation}
The ratio of the thermal ($S_\mathrm{thermal}$) and active ($S_\mathrm{active}$) force spectra can similarly be determined (see Supplementary Information).

Both ratios, extracted from the PSD shown in Fig.~\ref{fig:setup}\textbf{b}, are plotted in Fig.~\ref{fig:setup}\textbf{c}. From $G'/G''$ we see that the elastic modulus $G'$ dominates at low frequencies but that the ratio rapidly diminishes with increasing frequency, tending towards $G'/G''(\infty)=\cot{(\pi\alpha/2)}=0.52 \pm 0.01$ at infinity. This is consistent with the previously observed viscoelastic cellular response of living cells~\cite{rigato2017high,hoffman2006consensus}. The viscoelastic crossover frequency $f_G$, at which
the cell switches from elastic to viscous behaviour, is 320~kHz, also consistent with the  literature \cite{rigato2017high}.
From $S_\mathrm{active}/S_\mathrm{thermal}$ we can see that the active motion dominates the power spectrum by more than six orders of magnitude at low frequencies, but decays more rapidly with frequency than thermal noise, with thermal motion dominating above $f_S=46$~kHz (Fig.~\ref{fig:setup}\textbf{b} inset).

\begin{figure}[ht!]
\centering\includegraphics[width=\textwidth]{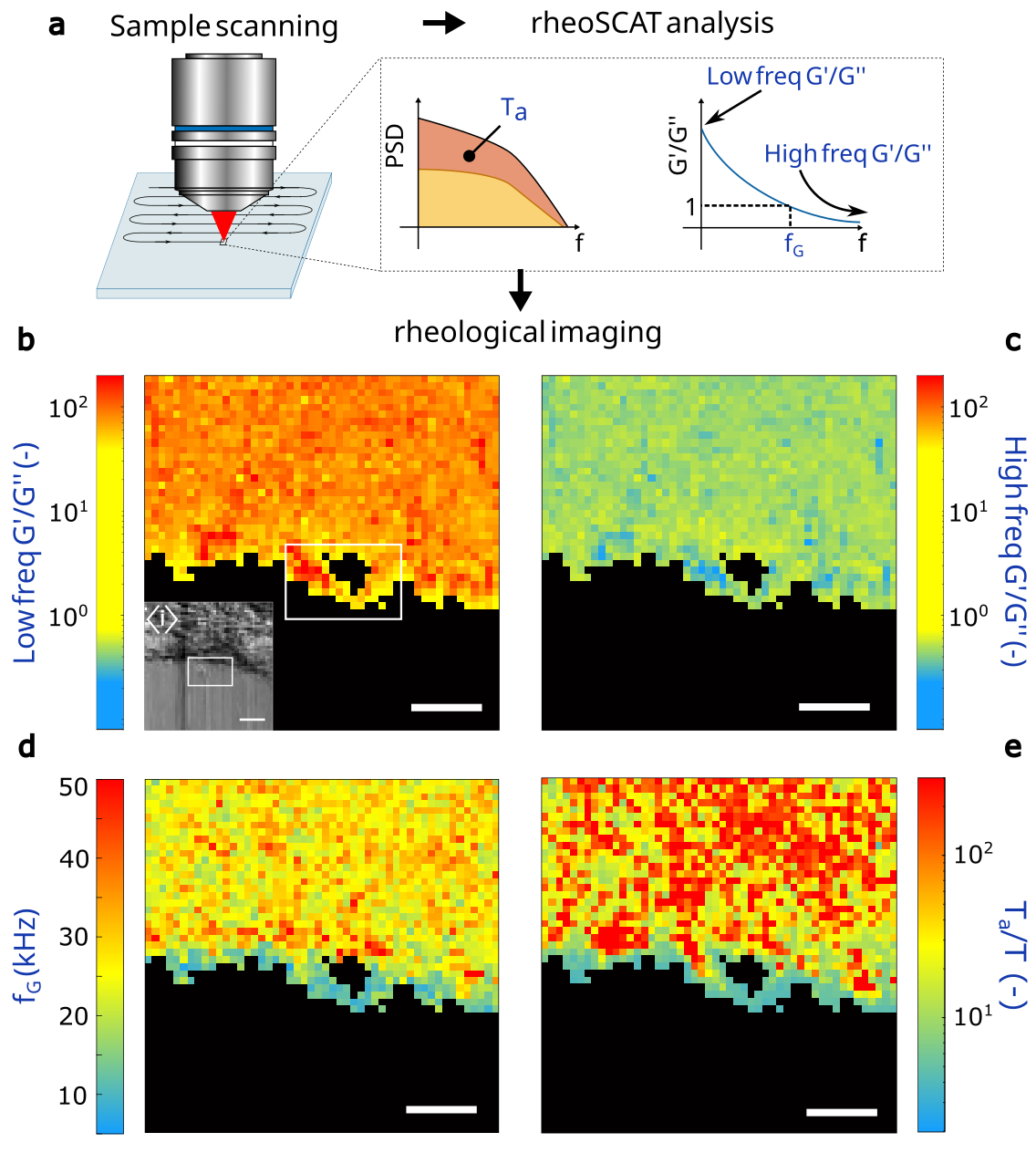}
\caption{
\textbf{rheoSCAT imaging.} \textbf{a} Imaging procedure: The sample is raster-scanned to acquire an intensity time trace at each pixel. Fourier analysis and model fitting are then applied to extract local rheological parameters, yielding spatially resolved rheoSCAT images. \textbf{b} rheoSCAT image of the ratio $G'/G''$ at low frequency (20~Hz). The inset corresponds to the mean intensity image calculated from the same data. \textbf{c} rheoSCAT image of the ratio $G'/G''$ at high frequency (infinity). \textbf{d} rheoSCAT image of the viscoelastic crossover frequency $f_G$. \textbf{e} rheoSCAT image of the effective active temperature normalised by the ambient temperature $T_a/T$. The Scale bars represent 5~$\mu$m.
}
\label{fig:HeLa}
\end{figure}

\subsection*{Spatially resolved rheological imaging}

The cellular rheological properties extracted from the rheoSCAT microscope can be spatially mapped to generate quantitative images. This is achieved by raster-scanning the sample in two dimensions using a high-precision nanostage and recording the rheoSCAT signal for 50~ms, acquiring 50$\times$50 pixel images over a 25$\mu$m$\times$25$\mu$m field of view in under 2.5 minutes (see Fig.~\ref{fig:HeLa}\textbf{a}).

We apply this method first to image a region of a HeLa cell. Since the model $S(f)$ is valid only within the cell, intracellular pixels were identified by integrating the PSD across frequency (see Supplementary Information), yielding a value connected to the mechanical energy within each voxel. Voxels inside the cell exhibited substantially higher energy, providing clear contrast between intracellular and extracellular regions (see Supplementary Information), consistent with enhanced motion and particle density inside the cytoplasm. Pixels with mechanical energy at least twice as large as the average energy in the background were selected as intracellular pixels.

Rheological images of an HeLa cell are displayed in Fig.~\ref{fig:HeLa}. Fig.~\ref{fig:HeLa}\textbf{b} and \textbf{c}, show the viscoelastic ratio $G'/G''$ at low (20~Hz, the lowest observable frequency within 50~ms acquisition time) and high (extrapolated to infinity) frequencies, respectively; Fig.~\ref{fig:HeLa}\textbf{d} shows the viscoelastic cross-over frequency, $f_G$; and Fig.~\ref{fig:HeLa}\textbf{e} shows the effective active temperature $T_a$.
rheoSCAT images of parameters $\kappa$, $\xi$ and $\alpha$ can also be found in the Supplementary Information.

The images of $G'/G''$ (Fig.~\ref{fig:HeLa}\textbf{b} and \textbf{c}) show that the viscoelastic ratio is fairly uniform across the cell except in close proximity to the cell edge. We find that elasticity is dominant at low frequencies ($G'/G''\gg 1$), with average $G'/G''$ values of 87 deep within the cell and 71  close to the cell boundary (within a few micron).
At high frequencies $G'/G'' < 1$, such that viscosity dominates, with an average value of 0.46.
Throughout the paper, we find that the uncertainties, given by the standard error across the sampled pixels, are very small -- in the examples given here, around 0.01~\% of the mean values -- and therefore omitted.
At high frequencies, $G'/G''$ is well within literature range  \cite{rigato2017high,hoffman2006consensus}. At low frequencies however, it is larger than what is observed with beads both within the cell interior and at cell boundary. This might be due to that fact that the endogenous rheoSCAT probe particles are better coupled to the cytoskeleton compared to beads that are typically trapped in vesicles inside the cell -- a limitation which has been identified as a serious issue for optical tweezers-based rheology \cite{wu2012high}.

The inset of Fig.~\ref{fig:HeLa}\textbf{b} shows the mean intensity image, obtained from the same data by averaging the photocurrent over the pixel dwell time (50 ms), analogous to DC-based interferometric imaging. 
Notably, the rheoSCAT images reveal fine structures that are not visible from the mean intensity (highlighted by white squares in Fig.~\ref{fig:HeLa}\textbf{b} and inset). These structures have dimensions and shape characteristic of filopodia, commonly found on HeLa cell membranes~\cite{bohil2006myosin}. 
This demonstrates that, by observing endogenous fluctuations, rheoSCAT microscopy can image structural features invisible in conventional DC interference images, extending beyond previous iSCAT measurements that have observed filopodia using labels \cite{taylor2019interferometric} or large endogenous  particles \cite{park2024label}. It also allows us to image the difference in viscoelasticity between the filopodia and other parts of the cell, an entirely new capability. Inside filopodia, we observe a higher elastic-to-viscous ratio at low frequencies (mean 120) and a lower ratio at high frequencies (mean 0.36) compared to the rest of the cell.
At low frequencies, the increased $G'/G''$ reflects the long-time elastic response of filopodia, which consist of prestressed, bundled actin filaments that can behave as stiff viscoelastic rods relative to the bulk cytoplasm~\cite{bornschlogl2013filopodial}.
At high frequencies, dissipation is dominated by frictional coupling between the actin bundle and the surrounding cortical network~\cite{bornschlogl2013filopodial} which can lead to enhanced viscous losses and a reduced $G'/G''$ ratio compared to the cell interior.

Figure~\ref{fig:HeLa}\textbf{d} shows the transition frequency $f_G$,  where the cell shifts from solid-like to fluid-like behaviour. $f_G$ is fairly uniform in the cell interior with a mean of 22~kHz. A distinct reduction is observed at the cell edge and inside the filopodia where $f_G$ is reduced to an average of 14~kHz. In both regions, the values of $f_G$ are consistent with the literature~\cite{rigato2017high}.

Imaging the effective temperature (Fig.~\ref{fig:HeLa}\textbf{e}) reveals significant variations across different regions of the cell. At the cell edge, pixels exhibited values with $T_a/T\sim5$ consistent with our previous observations. In contrast, the cytoplasmic interior reveals micron-scale heterogeneity with $T_a/T\sim20$ and $T_a/T\sim200$. While the fit uncertainty on these pixel is high, on average ${T_a}^{+40\%}_{-76\%}$, the spatial distribution of $T_a$ is not random as would be expected from statistical errors, but rather shows micron size clusters. This indicates heterogeneous active mechanics within the bulk cytoplasm. 
Such differences in effective temperature could originate from vesicle-rich regions, motor-driven regions (increasing $T_a$) or denser or more cross-linked actin regions (decreasing $T_a$) which have been observed to have micron scale length \cite{korabel2021local,xi2016molecular}.
Together, these results suggest that the cytoplasm is mechanically compartmentalized into distinct active states rather than behaving as a homogeneous active fluid.

\begin{figure}[ht!]
\centering\includegraphics[width=\textwidth]{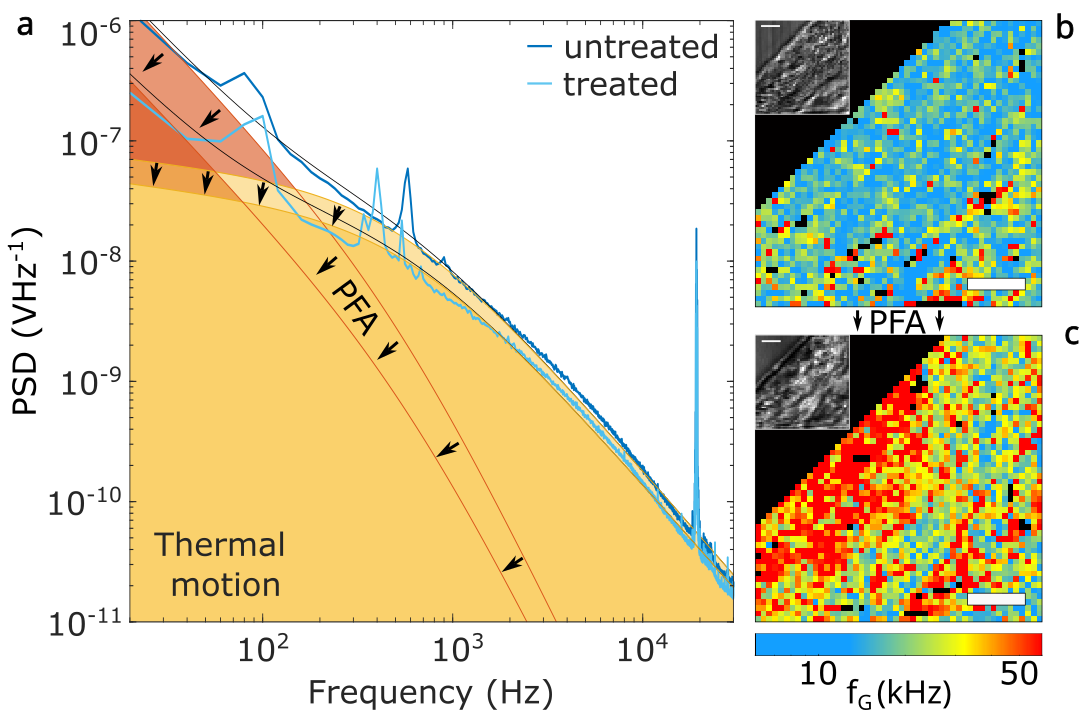}
\caption{\textbf{rheoSCAT parameters evolution.} 
\textbf{a} Averaged PSD over pixels of an A549 cell before (blue) and after (cyan) 10 minutes of PFA treatment. Both PSDs are fitted with the model $S(f)$ (black curves) and their thermal and active motion contribution are displayed in red and yellow shadings respectively. \textbf{b} and \textbf{c} rheoSCAT image of the viscoelastic crossover frequency $f_G$ before and after PFA treatment respectively. The insets show the corresponding mean intensity images. Scale bars represent \(5~\mu\text{m}\). 
}
\label{fig:A549}
\end{figure}

\subsection*{Imaging microrheological changes}

We next assess the ability of rheoSCAT microscopy to track time-dependent changes in cellular microrheology. To this end, we imaged A549 lung carcinoma epithelial cells, both before and after 10~minutes of exposure to 2\% paraformaldehyde (PFA; see Methods). In addition, a different A549 cells from the same sample was imaged 24~hours post-treatment (see Supplementary Information). PFA is a chemical fixative that stabilizes cellular structures by forming covalent cross-links between amino-acid side chains, increasing membrane permeability and ultimately inducing cell death~\cite{jamur2009cell}.

Figure~\ref{fig:A549}\textbf{a} shows the power spectral density (PSD) averaged over 2,000 intracellular pixels before (blue) and after (cyan) PFA exposure. Fits to the model $S(f)$ (black lines) reveal a 55\% reduction in total motion (thermal plus active) following fixation assault. Notably, the active contribution is reduced by 77\%, compared with a 33\% reduction in the thermal component. These trends are consistent with the effects of PFA: covalent cross-linking restricts the motion of cellular constituents, reducing thermal fluctuations, while immobilization of molecular motors and cytoskeletal stiffening strongly suppress active dynamics. After 24~hours, the PSDs exhibit a single power-law slope (see Supplementary Information), characteristic of purely thermal motion. This is expected following cell death, when active processes are fully suppressed and residual fluctuations arise from unbound components diffusing within the fixed cellular scaffold.

The average low frequency viscoelastic ratio, extracted from the PSDs fits, decreases from 31 to 25 after treatment, while at high frequencies it increases from 0.7 to 0.8 (see Supplementary Information). The increase at high frequencies is consistent with PFA-induced stiffening, which enhances the elastic response at short timescales \cite{kim2017mechanical}. Conversely, the reduction at low frequencies reflects the suppression of active prestress and cytoskeletal remodelling, which normally contribute to long-timescale elasticity \cite{gupta2016single}. After 24~hours, $G'/G''$ becomes frequency-independent with an average value of 0.5, consistent with the viscoelastic response of a dead cell.

Figures~\ref{fig:A549}\textbf{b} and \textbf{c} show rheoSCAT maps of the crossover frequency $f_G$ before and after PFA treatment. Before fixation, $f_G$ is relatively uniform across the cell, with a mean value of 18~kHz. After treatment, the average $f_G$ increases markedly to 43~kHz, and exhibits large-scale spatial structures with the largest changes occurring within the first 6–8~$\mu$m of the cell edge. This upward shift of the elastic-to-viscous transition indicates increased rigidity and slower stress relaxation, consistent with the observed changes in $G'/G''$. In contrast, the corresponding mean intensity images (insets) show only minor differences.

Finally, although the average active temperature $T_a$ decreases by approximately a factor of three following PFA treatment, spatially resolved $T_a$ maps reveal localized patches of elevated apparent activity (5–8~$\mu$m in size) near the cell periphery at intermediate treatment times (see Supplementary Information). After 24~hours, $T_a$ falls to zero within the accessible frequency range, as expected in the absence of active motion. The transient increases in $T_a$ observed shortly after PFA treatment are likely associated with cytoskeletal disruption and structural relaxation induced by PFA, rather than genuine motor-driven activity.

\section*{Discussion}

Label-free imaging techniques such as iSCAT and ROCS have recently begun to be used to measure cellular dynamics~\cite{junger2022100,hsiao2022spinning,trelin2024chiscat}. By achieving a twenty-fold increase in measurement bandwidth~\cite{hsiao2024probing}, and thereby resolving both the cytoskeleton’s elastic-to-viscous crossover and the transition from active to thermal force dominated dynamics, rheoSCAT connects these dynamics to microrheology.
This is significant because viscoelasticity modulates cell health and function~\cite{mofrad2009rheology}, yet has traditionally only been measurable either by using spatially coarse-grained methods such as optical tweezers~\cite{chaubet2020dynamic,rigato2017high,crocker2000two}, multi-particle tracking experiments that follow the independent trajectories of an ensemble of fluorescent labels~\cite{kole2004intracellular} and nanoparticles~\cite{guo2014probing}, or by using Brillouin microscopy, which measures compressional viscoelasticity at gigahertz frequencies and infers low-frequency cellular behaviour~\cite{keshmiri2024brillouin}.
In contrast, rheoSCAT enables direct, unlabeled imaging of viscoelasticity over large regions of live cells at frequencies relevant to biology, simultaneously capturing both mechanical properties and activity levels. This allows rheological mapping with higher spatial resolution, faster acquisition, and simpler instrumentation than existing techniques.
 
Whereas conventional microrheology typically relies on externally introduced probe particles~\cite{wu2012high,guo2014probing}, rheoSCAT exploits scattering from endogenous particles within the cytoplasm. It yields power spectral densities that are consistent with previous microrheology measurements~\cite{rigato2017high,chaubet2020dynamic,hoffman2006consensus}, fitting parameters consistent with the microrheology literature~\cite{rigato2017high,chaubet2020dynamic,hoffman2006consensus}, and fixation-induced changes that align with expectations.
A key question is why the complex, heterogeneous ensemble of particles in the cytoplasm can yield results consistent with established rheological measurements. The likely explanation lies in correlated motion among particles, which is known to occur across length scales comparable to the size of our microscope voxel~\cite{crocker2000two,crocker2007multiple}.

Since the intensity of dipole scattering scales as the sixth power of particle radius, one might expect large structures like mitochondria to dominate the rheoSCAT signal. However, we observe rheoSCAT spectra consistently at all pixels within the sampled cell. This implies that the dominant scattering elements are abundant and spatially continuous -- excluding sparse, larger structures as the primary contributors. Instead, the signal must arise from smaller, higher abundance and more weakly scattering particles. The existence of correlated motions explains how this can occur. Particles in the cytoplasm undergo partially coherent motion driven by cytoskeletal dynamics, superimposed with random local fluctuations~\cite{crocker2000two,crocker2007multiple}. The rheoSCAT power spectrum includes contributions from both types of motion, correlated and uncorrelated. Crucially, the relative contribution from correlated motions increases quadratically with particle number due to the presence of cross-correlation terms that are absent for purely random motions (see Supplementary Information). As a result, our estimates show that, due to their overwhelming abundance, correlated motion from proteins and protein complexes is likely the primary signal source (see Supplementary Information). That the microscope is sensitive to protein motions is consistent  with iSCAT’s demonstrated ability to detect individual proteins~\cite{piliarik2014direct} and to observe fluctuations due to their motions within the nucleus~\cite{mazaheri2024iscat}. The dominance of correlated protein motions, driven by cytoskeleton dynamics, indicates that rheoSCAT measurements probe the bulk viscoelasticity across the microcope voxel rather than local diffusive motion, explaining the agreement with models of cytoskeletal viscoelasticity.
This new understanding  may also help explain the biophysical basis of activity signals previously observed with ROCS~\cite{junger2022100} and iSCAT~\cite{hsiao2024probing}.

We have shown that rheoSCAT can extract detailed, spatially resolved information about the material properties and activity of cells, including overall activity levels, diffusion and the ratio of storage and loss moduli as a function of frequency ($G'(f)/G''(f)$). An outstanding question is whether it is possible to achieve an absolute calibration of the complex shear modulus. Typically, such calibration requires detailed knowledge of the size and shape of the probe particles~\cite{brau2007passive}. It is conceivable that the ensemble of cytoplasmic proteins that contribute to the rheoSCAT signal is sufficiently uniform, within a given cell type, to permit such a calibration. Alternatively, in two-particle microrheology correlated particle motions allow calibration without detailed knowledge of the probe particles~\cite{levine2000one}. Similar techniques may be applicable to ensembles of correlated particles such as those in rheoSCAT. These remain open questions for future investigation.

A limitation of our current rheoSCAT microscope is its slow acquisition time (2.5 minutes per frame). If only activity above a few kilohertz was of interest, the pixel dwell time could be reduced to beneath a millisecond using two-dimensional acousto-optic modulators (AOM) or resonant scanning~\cite{terrasson2024fast}. This would reduce the acquisition time for 50$\times$50 pixel images to two seconds. Video rate imaging could be achieved using a wide-field configuration and high-speed camera. This would allow the observation of spacial dynamics down to 20~Hz at the expense of reduced signal bandwidth.

\section*{Methods}

\subsection*{\textbf{Cell culture conditions}}
HeLa (CCL-2) and A549 (CCL-185) were maintained in Dulbecco's Modified Eagle Medium (DMEM) supplement with 10\% foetal bovine serum (FBS),  2mM L-glutamine and 100 units/ml penicillin/streptomycin in a humidified 37 degree incubator supplement with 5\% CO2.

\subsection*{\textbf{PSD fits}}
To improve the accuracy and reproducibility of the PSD fits, each pixel’s PSD is normalised by its mean intensity and corrected by subtracting the shot-noise contribution. Fits are performed on a logarithmic scale, and frequency points around the noise peaks at 19 and 38~kHz are excluded.

\subsection*{\textbf{Cell fixation}}
To conduct the cell fixation experiment, the sample medium was carefully partially removed using a pipette, ensuring that the microscope alignment remained undisturbed. The medium was then replaced with a fresh solution supplemented with 2\% PFA. Imaging was resumed 10 minutes after the introduction of the PFA solution to allow sufficient time for the fixation process to initiate.
Following initial measurements, the sample was removed from the microscope and stored overnight in a sterile environment at room temperature. After twenty four hours of PFA treatment, the same sample was re-mounted and re-imaged to assess the long-term fixation effects.

\begin{backmatter}
\bmsection{Funding}
This work was primarily supported by the Air Force Office of Scientific Research (grant no. FA9550-22-1-0047 and FA9550-24-1-0286). It was also supported by the Australian Research Council Centre of Excellence in Quantum Biotechnology (grant no. CE230100021).

\bmsection{Acknowledgments}

Some of the cells used in this research were derived from a HeLa cell line. Henrietta Lacks, and the HeLa cell line that was established from her tumor cells without her knowledge or consent in 1951, have made significant contributions to scientific progress and advances in human health. We are grateful to Henrietta Lacks, now deceased, and to her surviving family members for their contributions to research.

The author acknowledges HC Photonics for the discount on the second harmonic generator module, which contains the periodically poled nonlinear waveguide used for converting 1560 nm light to 780 nm.

\bmsection{Ethics declarations}
The authors declare no conflicts of interest.

\bmsection{Data Availability Statement}
The data supporting the findings of this study are available within the paper and its Supplementary Information.  All other data that support the findings of this study are available from the corresponding author on request.

\bmsection{Supplemental document}
See Supplement for supporting content. 

\end{backmatter}


\bibliography{main}

\end{document}